\documentclass[aps,amsmath,amssymb,pr,twocolumn]{revtex4-1}
\usepackage{bm}
\usepackage{epsfig}
\usepackage{amsmath}
\usepackage{color}
\newcommand{\nix}[1]{}
\usepackage[unicode=true,colorlinks=true,citecolor=magenta]{hyperref}

\begin{document}

\title{Circular photocurrent in Weyl semimetals with mirror symmetry}

\author{N.\,V.\,Leppenen, E.\,L.\,Ivchenko, and L.\,E.\,Golub}
\affiliation{Ioffe Institute, St.~Petersburg 194021, Russia}

%\date{\today}

\begin{abstract}
We have studied theoretically the Weyl semimetals the point symmetry group of which has reflection planes and which contain equivalent valleys with opposite chiralities. These include the most frequently studied compounds, namely the  transition metals monopnictides TaAs, NbAs, TaP, NbP, and also Bi$_{1-x}$Sb$_x$ alloys. The circular photogalvanic current, which inverts its direction under reversal of the light circular polarization, has been calculated for the light absorption under direct optical transitions near the Weyl points. In the studied materials, the total contribution of all the valleys to the photocurrent is nonzero only beyond the simple Weyl model, namely, if the effective electron Hamiltonian is extended to contain either an anisotropic spin-dependent linear contribution together with a spin-independent tilt or a spin-dependent contribution cubic in the electron wave vector ${\bm k}$. With allowance for the tilt of the energy dispersion cone in a Weyl semimetal of the $C_{4v}$ symmetry, the photogalvanic current is expressed in terms of the components of the second-rank symmetric tensor that determines the energy spectrum of the carriers near the Weyl node; at low temperature, this contribution to the photocurrent is generated within a certain limited frequency range $\Delta $. The photocurrent due to the cubic corrections, in the optical absorption region, is proportional to the light frequency squared and generated both inside and outside the $\Delta$ window.
\end{abstract}

\maketitle

\section{Introduction}
Recently, three-dimensional systems with a linear spectrum called Weyl's semimetals have been discovered and are actively studied; see \cite{RMP_2018,Hasan2018,Kar2019} for recent reviews.
These systems which have a nondegenerate energy spectrum of quasi-particles (with the exception of double degeneracy in the Weyl node) attract much attention due to their unusual electrical and optical properties in the bulk and on the surface and expand the concepts of topological theory in solid state physics. In  the simplest model of such systems, the current carriers are described by the effective Hamiltonian which has the form of the Weyl Hamiltonian used to describe neutrinos, which explains their name.

In Weyl's semimetals, the ``circular'' photocurrent, i.e., an electric current arising under the light absorption at zero bias and reversing its direction to the opposite when  the light circular polarization changes its sign, behaves in a remarkable way. Namely, it has been shown that in the absence of symmetry (mirror reflection) planes, the circular photocurrent is directed along the photon momentum and its generation rate is determined, in addition to the electric field strength of the electromagnetic wave, by the world constants ~\cite{Moore}.

Real Weyl semimetals TaAs, TaP, NbAs, NbP \cite{TaAsdispersion, TaNbAsP, TaNbAsP2, TaAsPoints, TaAspreasure, Gedik, TaAs2018} and Bi$_ {1-x}$Sb$_x$~\cite {BiSb} have point symmetries C$_{4v}$ and C$_ {3v}$ respectively. Thus, they have a mirror symmetry in which case the contributions to the circular photocurrent from two Weyl nodes connected by the reflection compensate each other exactly. As a result, there is no  electric current and only a pure valley photocurrent is generated in the direction of light propagation. However, the above-mentioned point symmetry groups are gyrotropic so that a circular electric photocurrent transverse to the photon momentum is possible in the relevant materials when the light propagates perpendicularly to the 3rd or 4th rotary axis of symmetry. Such a photocurrent cannot be microscopically obtained within the framework of the pure Weyl Hamiltonian. It has been shown in Ref.~\cite{Patrick} that the transverse circular photocurrent can appear due to spin-independent  corrections  to the Weyl Hamiltonian linearly dependent on the electron wave vector and leading to ``tilt'' of the Weyl cones. Within the framework of such a model, an attempt has been made to explain the experimental results on the circular photogalvanic effect in tantalum arsenide observed under CO$_2$ laser excitation with a photon energy of 120~meV~\cite{Gedik}.

In this paper we demonstrate that the tilt of the electron dispersion leads to a photocurrent only within in a limited frequency range. We propose an alternative model based on nonlinear spin-dependent corrections to the Weyl Hamiltonian and leading to a photocurrent that increases monotonously with the increasing light frequency. The theory is developed for an arbitrary matrix $\beta_{ij}$ describing the linear part of the effective electron Hamiltonian $\beta_ {ij} \sigma_i k_j$ and containing both diagonal and off-diagonal components.
\section{Effective Hamiltonian}
Near the Weyl point $ {\bm q}_W$, it suffices to take into account only two bands touching each other at the Weyl point. In general, the effective Hamiltonian is a 2nd rank matrix which can be represented in the following form
\begin{equation} \label{Weyl2}
{\cal H}({\bm q}) =  {\bm \sigma}\cdot {\bm d}({\bm q}) + \sigma_0 d_0({\bm q})\:, 
\end{equation}
where $\bm q$ is the wave vector defined in the first Brillouin zone of the crystal, $\sigma_i$ are the Pauli spin matrices, $\sigma_0$ is the unity 2$\times$2 matrix, and at ${\bm q}= {\bm q}_W$ the three-component function ${\bm d}({\bm q})$ and the scalar function $d_0({\bm q})$ vanish. For the energy eigenvalues, one has
\begin{equation} \label{Energy}
E_{\pm, {\bm q}} =  \pm d({\bm q}) + d_0({\bm q})\:,
\end{equation} 
where $d(\bm q) = \sqrt{|{\bm d}(\bm q) |^2}\:. $ 

\subsection{The linear approximation with anisotropic spectrum}
In the linear approximation with respect to the deviation of the wave vector ${\bm q}$ from ${\bm q}_W $ the effective Hamiltonian 
\begin{equation} \label{linearH}
{\cal H} = \beta_{ij} \sigma_i \left( q_j - q_{Wj} \right) + \sigma_0 a_l \left( q_l - q_{Wl} \right)\:,
\end{equation}
with $i,j,l$ being the Cartesian coordinates $x,y,z$, is determined by nine components of the 3$\times$3 matrix $\hat{\bm \beta}$ and three components of the vector ${\bm a}$. They are found by differentiating of $d_i$ and $d_0$ with respect to ${\bm q}$
\begin{equation}
\beta_{ij}= \left( \frac{\partial d_i}{\partial q_j}\right)_{{\bm q}_W}\:,\quad a_l = \left( \frac{\partial d_0}{\partial q_l}\right)_{{\bm q}_W}\:.
\end{equation}
We remind that the vector ${\bm a}$ in the spin-independent term in~(\ref{linearH}) is named as the tilt.

In what follows, we denote the difference ${\bm q} - {\bm q}_W $ as ${\bm k} $, i.e., we count the electron wave vector from the Weyl node. Then, in the linear approximation, Eq.~(\ref{Energy}) can be represented as
\begin{equation} \label{Energy2}
E_{\pm, {\bm k}} = \pm \sqrt{\Lambda_{ij} k_i k_j} + {\bm a} \cdot {\bm k} \:.
\end{equation}
Here, the symmetric positively-definite matrix $\hat{\bm \Lambda} $ is related to the matrix $\hat {\bm \beta} $ by
\begin{equation} \label{Lambda}
\hat{\bm \Lambda} = \hat{\bm \beta}^T \hat{\bm \beta} \:,
\end{equation} 
where $\hat{\bm \beta}^T$  is the matrix transposed with respect to $\hat{\bm \beta}$.
To find six linearly independent components of the $\hat{\bm \Lambda}$ matrix, it suffices to calculate the linear dispersion near the Weyl point for six different directions in the ${\bm k}$-space.

For a given $\hat{\bm \Lambda}$ matrix, Eq.~(\ref{Lambda}) is satisfied by the matrices of the set
\begin{equation} \label{betaD}
\hat{\bm \beta} = \pm \hat{\bm D}(\varphi,\theta,\psi) \hat{\bm \beta}_0\:,
\end{equation}
where $\hat{\bm \beta}_0$ is any particular solution of this equation, and $\hat {\bm D}$  is the  matrix of orthogonal transformation of Pauli spin matrices $\sigma_i$, it is determined by three Euler angles $\varphi$, $\theta$ and $\psi$ and has the determinant $+1$. Indeed, the product $\hat{\bm \beta}^T \hat{\bm \beta}$ is equal to $\hat{\bm \beta}_0^T \hat{\bm D}^T \hat{\bm D} \hat{\bm \beta}_0 = \hat{\bm \beta}_0^T \hat{\bm \beta}_0$, and this, by the definition of $\hat{\bm \beta}_0$, is the $\hat{\bm \Lambda}$ matrix. Appendix~\ref{App1} specifies the method for selecting the matrix $\hat {\bm \beta}_0 $ and provides an equation for it expressed in terms of the matrix $\hat {\bm \Lambda} $. Note that the sign in~\eqref {betaD} is not determined by the matrix $\hat {\bm \Lambda} $ and represents the topological charge, or chirality, of this node \cite{RMP_2018, GolubIvch}
\begin{equation} \label{chirality}
{\cal C} = {\rm sgn} \{{\rm det}(\hat{\bm \beta})\}\:.
\end{equation}
The Weyl nodes related by an improper symmetry operation $\sigma_v$, for example, the nodes ${\bm k}_{W1} = (k_x, k_y, k_z) $ and $ {\bm k}_{W2} = (-k_x, k_y, k_z)$, have opposite chiralities. At the same time, for the nodes connected by the time inversion operation $ {\cal T}$, the $\hat {\bm \beta} $ matrices coincide, and therefore their chiralities also coincide. As it should be for a real matrix of dimension 3$\times$3, with the $\pm$ sign selected in Eq.~(\ref{betaD}), the set of matrices $\hat {\bm \beta} $ is determined by nine parameters: the six components of the symmetric matrix $\hat{\bm \Lambda}$ and three angles $\varphi, \theta, \psi$.

The transformation $\hat{\bm D}$ in Eq.~(\ref{betaD}) changes one basis of a doubly degenerate state at the Weyl node to another while maintaining the coordinate system in the $ {\bm k}$-space. Naturally, the electron energy at the point ${\bm k}$ does not change. A similar  transformation $\hat {\bm D}$ in the spin space, $\sigma_i = D^{- 1}_ {ii'} \sigma_{i'}$, can also be applied to the general Hamiltonian~(\ref{Weyl2}). In this case, the scalar product $\sigma_i d_i ({\bm k})$ transfers to the sum $ \sigma_{i'} d_{i'}({\bm k}) $, where $ d_{i'} = D_{i 'i} d_i$. For further calculation of the photocurrent, it is important to bear in mind that the above transformation, which does not involve the ${\bm k}$-space, does not change not only the electron energy but also the Berry curvature ${\bm \Omega}_{\bm k}$. For proof, we note that the components of the latter can be represented as a scalar-vector product \cite{GoIvSp, GolubIvch}
\begin{equation} \label{Berrygeneral}
\Omega_{{\bm k},i} = \frac{\bm d }{2d^3} \cdot  
\left( \frac{\partial \bm d }{\partial k_{i+1}} \times \frac{\partial \bm d}{\partial k_{i+2}} \right) \:,
\end{equation}
which is invariant with respect to any rotation of the vector ${\bm d}({\bm k})$.

A Weyl semimetal with the Hamiltonian 
\begin{equation} \label{betaa}
{\cal H} = \beta {\bm \sigma}\cdot{\bm k} + {\bm a}\cdot {\bm k}
\end{equation}
belongs to type I if $|\beta| > |{\bm a}|\equiv a$, and to type II if $|\beta| < a$ \cite{typeII}. For a general form of the matrix $\beta_{ij}$, the membership to type I or II is determined respectively by the absence or presence of such a direction of the vector ${\bm k}$ where $d({\bm k}) < | {\bm a}\cdot {\bm k}|$, or
\begin{equation} \label{types}
\sum\limits_{ij}\Lambda_{ij} k_i k_j < \sum\limits_{ij} a_i a_j k_i k_j\:.
\end{equation}
The criterion for fulfillment of this condition is the inequality
\begin{equation} \label{types2}
1 < |{\bm b}| \equiv b\:,
\end{equation}
where ${\bm b}$ is a vector with the components 
\begin{equation} \label{tildea}
b_i = \beta^{-1}_{ji}a_j
\end{equation}
and the absolute value $b = \sqrt{\Lambda^{-1}_{ij} a_ia_j}$. In this paper we consider the type-I Weyl semimetals ($b<1$).

The table S3 in Supplementary to the article \cite{Gedik} contains calculated values $v_{\pm}(lmn) = \partial E_{\pm,{\bm k}}/\partial k$ for the six directions $[lmn]$ in the ${\bm k}$-space of the TaAs semimetal:  [100], [010], [001], [110], [101] and [011]. By using these six values, we have calculated the $\hat{\bm \Lambda}$ matrix and the vector ${\bm a}$ entering  Eq.~(\ref{Energy2}). The determinant of the obtained matrix $\hat{\bm \Lambda}$ turned out to be negative in contradiction with the fact that $\hat{\bm \Lambda}$ must be a positive definite matrix. If we assign the values of $v_{\pm}(101)$ to the direction $ [10 \bar{1}] $ rather than to the direction [101], the matrix $\hat {\bm \Lambda}$ becomes positively defined. However, the absolute value of the vector ${\bm b}$ exceeds unity, which is inconsistent with the identification of TaAs as a type-I semimetal \cite{Hasan2018}. Therefore, the table S3 requires clarification.
\subsection{Allowance for {\bf k}-nonlinear terms}
When calculating the photocurrents in the Weyl semimetals in Sec.~\ref{Sect. V}, we take into account nonlinear in ${\bm k}$ terms in the electron effective Hamiltonian
\begin{equation} \label{linnonlin}
{\cal H} = \beta_{ij} \sigma_i k_j + \sigma_i P^{(2)}_i({\bm k}) + \sigma_i P^{(3)}_i({\bm k})
+ \sigma_0 \left({\bm a}\cdot {\bm k} +\dots \right),
\end{equation}
where for generality we included both quadratic and cubic spin-dependent contributions,
\begin{align} \label{23}
P^{(2)}_i({\bm k}) = \sum\limits_{jl} C_{ijl}^{(2)}k_jk_l\:, \nonumber \\ P^{(3)}_i({\bm k}) = \sum\limits_{jlm}C^{(3)}_{ijlm}k_jk_lk_m\:.
\end{align}
With allowance for the nonlinear terms, the energy can still be represented by Eq.~(\ref{Energy}) but the vector ${\bm d} ({\bm k})$ should be generalized  as
\[
d_i({\bm k}) = \beta_{ij} k_j + P^{(2)}_i({\bm k}) +  P^{(3)}_i({\bm k})\:.
\]

The coefficients introduced in Eq.~(\ref {23}) are symmetric with respect to the permutations of the indices $jl$ and $jlm$, respectively,
\[
C_{ijl}^{(2)} = C_{ilj}^{(2)}\:, \qquad C^{(3)}_{ijlm} = C^{(3)}_{iljm} = C^{(3)}_{imlj}\:.
\]
Under the $\hat{\bm D}(\varphi, \theta, \psi)$ orthogonal transformation, the matrices ${\bm C}^{(n)} $ transfer into the matrices
\[
\sum\limits_i D_{ni} C^{(2)}_{ijl}\hspace{3 mm}\mbox{and} \hspace{3 mm} \sum\limits_i D_{ni} C^{(3)}_{ijlm}\:.
\]
Such a transformation in Eq.~(\ref{betaD}) leaves invariant not only the matrix $\hat{\bm \Lambda}$ but also the sums of the products
\begin{eqnarray} \label{NonlinInv}
&&C_{jlmn} = \sum\limits_i \beta^{-1}_{ji} C^{(3)}_{ilmn}\:,\\
&&C'_{jlmn} = \sum\limits_i \beta_{ij} C^{(3)}_{ilmn} = \sum\limits_i \beta^T_{ji} C^{(3)}_{ilmn}\:.\nonumber
\end{eqnarray}
The matrices $\hat{\bm C}$ and $\hat{\bm C}'$ are related by 
\begin{equation} \label{SS'}
C'_{jlmn} = \Lambda_{ji}C_{ilmn}\:.
\end{equation}
\section{General equations for the photocurrent}
In continuation of the works \cite{GoIvSp, GolubIvch} we investigate the circular photogalvanic effect described by the phenomenological formula
\begin{equation} \label{CPGE}
j_{\lambda} = \gamma_{\lambda \eta} \varkappa_{\eta} |{\bm E}|^2\:,
\end{equation}
where $\bm j$ is the photocurrent density, ${\bm E}$ is the amplitude of the electric field of the light wave, ${\bm \varkappa} = i({\bm E} \times {\bm E}^*)/|{\bm E}|^2$ is the photon vector chirality 
equal to the unit vector in the direction of light propagation multiplied by its degree of circular polarization $P_{\rm circ}$. In gyrotropic crystal classes C$_{nv}$ ($n = 3,4,6$), the tensor ${\bm \gamma}$ has two nonzero components $\gamma_{xy}= - \gamma_{yx}$ the calculation of which is the purpose of this work.

The density of the circular photocurrent generated in Weyl semimetals under direct optical transitions is calculated according to the following formula~\cite{Moore, GoIvSp, GolubIvch}
\begin{align} \label{MOmega}
{\bm j} = |{\bm E}|^2 \frac{2\pi e^3 }{\hbar}  \sum_{\bm k} &{\bm v}_{{\rm s}\mbox{-}{\rm o}} \tau_p 
\left({\bm \varkappa}\cdot{\bm \Omega_{\bm k}}\right)
\delta [2 d({\bm k})  - \hbar \omega]\\ &\times
 \left[ f_0 (E_{-,{\bm k}}) - f_0 (E_{+,{\bm k}}) \right]   .
 \nonumber
\end{align}
Here $e$ is the electron charge, ${\bm v}_{{\rm s}\mbox{-}{\rm o}} = \hbar^{-1} (\partial d/ \partial {\bm k})$ $-$ is the spin-dependent part of the electron group velocity in the conduction band, $\tau_p$ is is the electron momentum relaxation time, ${\bm \Omega_{\bm k}}$ is the Berry curvature (\ref{Berrygeneral}), $f_0(E) =\{\exp{[(E - \mu)/T]} + 1\}^{-1}$ is the equilibrium distribution function over the energy $E$, $\mu$ is the chemical
potential, because of the energy conservation law one has 
\begin{equation} \label{energytilt}
E_{\pm,{\bm k}} = \pm \frac{\hbar \omega}{2} + d_0({\bm k})\:.
\end{equation}

Fernando~de~Juan et al. \cite{Moore} have calculated the circular photocurrent ~(\ref{MOmega}) generating in the vicinity of the Weyl node with the linear ${\bm k}$ effective Hamiltonian (\ref{linearH}) without a tilt (${\bm a} = $ 0) and shown that the contribution of each node to the circular photocurrent has a universal form 
\begin{equation} \label{beta00}
{\bm j} = {\cal C} \Gamma_0\tau_p {\bm \varkappa} |{\bm E}|^2\:,
\qquad \Gamma_0 = {\pi e^3 \over 3 h^2},
\end{equation}
where $h$ is thePlanck constant, and ${\cal C} = \pm 1$ is the chirality of this node defined by Eq.~(\ref{chirality}). If there are improper elements of the crystal point symmetry, the contributions from nodes of opposite chiralities compensate each other and the photocurrent calculated in the  model of Ref.~\cite {Moore} vanishes. In such crystals, the photocurrent summed over all the Weyl nodes is different from zero when taking into account the tilt~\cite {Patrick} or nonlinear terms in the Hamiltonian~(\ref {linnonlin})~\cite {GoIvSp, GolubIvch}.

For definiteness, we consider the Weyl nodes with $k_z \neq0$. 
It is convenient to distinguish them by the point-symmetry elements of $g$ and products ${\cal T} g$ of the space operation $g$ and the time reversal operation ${\cal T}$. The tensor ${\bm \gamma}$ summed over all the valleys is written as a sum
\[
{\bm \gamma} = 2 \sum_g {\bm \gamma}^{(g)}\:, 
\]
where ${\bm \gamma}^{(g)}$ is the contribution of the valley $g{\bm k}_W$,  ${\bm k}_W$ is the initial node, e.g., that with $k_{W,y}> k_{Wx} > 0$; the factor of 2 takes into account the contributions of the $-g {\bm k}_W$ valleys. By using the general equation~(\ref {MOmega}) we can express the partial contribution of the valley $g{\bm k}_W $ to the photogalvanic tensor in the form
\begin{align} \label{gMOmega}
\gamma_{\lambda\nu}^{(g)} =\frac{2\pi e^3 \tau_p}{\hbar}  \sum_{\bm k}& v^{(g)}_{{\rm s}\mbox{-}{\rm o},\lambda} \Omega_{ {\bm k}, \nu}^{(g)} 
 \delta [2 d^{(g)}({\bm k})  - \hbar \omega]\\ & \times 
\left[ f_0 ( E^{(g)}_{-,{\bm k}} ) - f_0 (E^{(g)}_{+,{\bm k}}) \right]  \:. \nonumber
\end{align}
The particular component $\gamma_{xy}$ is given by
\begin{align} \label{gammaxyg}
&\hspace{2 cm}\gamma_{xy}^{(g)} =\frac{2\pi e^3 \tau_p}{\hbar^2} \\ 
\times \sum_{\bm k}&
\frac{1}{2d^3(g{\bm k})}  \frac{\partial d(g{\bm k})}{\partial k_x}\left[ {\bm d}(g{\bm k}) \cdot \left( \frac{\partial {\bm d}(g{\bm k}) }{\partial k_z} \times \frac{\partial {\bm d}(g{\bm k})}{\partial k_x} \right)\right] \nonumber
\\ 
&\times
\left[ f_0 ( E_{-,{g\bm k}} ) - f_0 (E_{+,{g\bm k}}) \right]  \delta [2 d(g{\bm k})  - \hbar \omega]\: .\nonumber
\end{align}
Due to the energy conservation law the value of $2 d^3(g{\bm k})$ can be replaced by $(\hbar \omega)^3/4$.

Changing the variables ${\bm k} \to g^{-1} {\bm k}$ in the sum (\ref{gammaxyg}) and taking into account that for one half of the symmetry operations $(g{\bm k})_x = \pm k_x$, for the second half $(g \bm k)_x = \pm k_y$, and for all $g \in C_{4v}$, the component of the wave vector $k_z$ does not change, we obtain for the $xy$-component of the total tensor ${\bm \gamma}$
\begin{align} 
\label{gamma_tot}
&\hspace{2 cm}\gamma_{xy} =\frac{4N\pi e^3 \tau_p}{\hbar^2 (\hbar \omega)^3}  \\
&\times \sum_{\bm k}\left[ f_0 ( E_{-,{\bm k}} ) - f_0 (E_{+,{\bm k}}) \right]  \delta [2 d({\bm k})  - \hbar \omega]\nonumber
\\
&\times  
\left\{  \frac{\partial d({\bm k})}{\partial k_x}\left[ {\bm d}({\bm k}) \cdot \left( \frac{\partial {\bm d}({\bm k}) }{\partial k_z} \times \frac{\partial {\bm d}({\bm k})}{\partial k_x} \right)\right]
+ (x \to y) \right\} ,\nonumber
\end{align}
where $N=16$ is the number of Weyl valleys.
\section{Circular photocurrent with account for the tilt}
In this section, we leave only ${\bm k}$-linear spin-dependent and spin-independent terms in the Hamiltonian (\ref{linnonlin}).
In Ref.~\cite {Patrick}, the tilt term was taken into account in the model with an isotropic matrix $\hat{\bm \beta} = \beta {\bm \sigma}\cdot {\bm k}$. Here we generalize the theory to arbitrary matrix $\hat{\bm \beta}$.

The $\hat{\bm \beta}$  matrix enters into Eq.~\eqref{gamma_tot}
only through $\hat{\bm \Lambda}$ and ${\cal C}$:
\begin{equation}
\label{v_Omega_lin}
\frac{\partial d}{\partial k_x} = {(\hat{\bm \Lambda}\bm k)_x\over  d(\bm k)},
\quad  {\bm d} \cdot \left( \frac{\partial {\bm d} }{\partial k_z} \times \frac{\partial {\bm d}}{\partial k_x} \right) 
 = {\cal C} \sqrt{\Delta_{\Lambda}} k_y,
\end{equation}
where $\Delta_{\Lambda} = {\rm det}(\hat{\bm \Lambda})$. 
Using Eqs.~(\ref{v_Omega_lin}) we can convert
$\gamma_{xy}$ to 
\begin{align} \label{betafinal}
&\hspace{1.5 cm}\gamma_{xy} = 
 \frac{12{\cal C}N \Gamma_0 \tau_p\sqrt{\Delta_{\Lambda}}}{\pi(\hbar \omega)^4} 
\\ &
\times 
\int  \text{d}{\bm k} \sum_{j}
\left( \Lambda_{xj} k_y - \Lambda_{yj} k_x \right) k_j F({\bm a}\cdot{\bm k}) \  \delta [2 d({\bm k})  - \hbar \omega], \nonumber
\end{align}
where 
\begin{equation} \label{Fk}
F(x) =  f_0 \left( - \frac{\hbar \omega}{2} + x \right) - f_0 \left(  \frac{\hbar \omega}{2} + x \right)\:. 
\end{equation}
Turning to the variables
\begin{equation} \label{QLambda}
Q_i = \beta_{ij} k_j\:,\quad k_j = \beta^{-1}_{ji} Q_i\:, \quad d({\bm k}) = Q \equiv |{\bm Q}|
\end{equation}
and integrating over the absolute value of the vector $\bm Q$, we obtain
\begin{align}\label{gammamid}
\gamma_{xy} = \frac{3}{2} &	\mathcal{C}N\Gamma_0\tau_p \\ 
\times \sum\limits_{st}  \left(\beta_{sx}\beta_{yt}^{-1} - \beta_{sy}\beta_{xt}^{-1}\right) &
\left\langle\frac{Q_sQ_t}{Q^2}F({\bm b}\cdot{\bm Q})\right\rangle\:, \nonumber
\end{align}
where the vector ${\bm b}$ is defined according to Eq.~(\ref{tildea}), and the angle brackets mean averaging over the directions of the vector~${\bm Q}$. If the tilt is neglected, the factor $F({\bm a}\cdot{\bm k}) = F({\bm b}\cdot{\bm Q})$ in Eq.~{betafinal} is independent of ${\bm Q}$, see Eq.~(\ref{energytilt}), and since the average $\langle Q_{s}Q_t \rangle$ equals to $ \delta_{st} Q^2/3$, the sum over $s$ and $t$ vanishes.

The bracketed average in Eq.~(\ref{gammamid}) can be converted to
\begin{align}
\left\langle\frac{Q_sQ_t}{Q^2}F({\bm b}\cdot{\bm Q})\right\rangle = {\delta_{st} \over 4} \int\limits_{-1}^1 (1 - u^2) F(bQu) du \nonumber
\\ +\ \frac{b_s b_t}{4b^2}
 \int\limits_{-1}^1 (3u^2 - 1) F(bQu) d u \:. \hspace{1 cm}\nonumber
\end{align}
The first integral makes no contribution to $\gamma_{xy}$, while the contribution of the second integral is reduced to
\begin{equation} \label{int}
\gamma_{xy} = \frac{3}{8} \mathcal{C}N\Gamma_0\tau_p \chi_{xy} J \:,
\end{equation}
where
\begin{equation} \label{xixy}
\chi_{xy} = \frac{\sum\limits_i \left( a_x \Lambda^{-1}_{yi}  -  a_y \Lambda^{-1}_{xi} \right) a_i}{\sum\limits_{ij} \Lambda^{-1}_{ij} a_ia_j}
\end{equation}
and
\begin{equation} \label{Jxy}
J =\int\limits_{-1}^1 (3u^2 - 1) F(bQu) d u\:.
\end{equation}

For ${\bm a}$~$\neq$~$0$, the expression for $\chi_{xy}$ is invariant to the
transformation~(\ref {betaD}). For an isotropic matrix  $\Lambda_{ij} = \Lambda \delta_{ij}$, the sum over $i$ in the numerator of~Eq.~\eqref{xixy} vanishes, in agreement with Refs.~\cite{GoIvSp, GolubIvch}, where it is shown that, in crystals of the C$_{nv}$ symmetry $(n=3,4,6)$, a circular photocurrent is absent even for nonzero tilt $d_0({\bm k})$, provided the spin-dependent part of the Hamiltonian has a simple isotropic form~(\ref {betaa}). However, when the matrix $\beta_{ij} $ is of a general form, the presence of the tilt term in the distribution functions $f_0 (E_ {\pm, {\bm k}})$ leads to the appearance of a photocurrent. For example, for a diagonal matrix $\Lambda_{ij}$ with different components $\Lambda_ {xx}$ and $\Lambda_{yy}$, the parameter $\chi_{xy} $ is equal to the ratio
$$
\frac{a_xa_y(\Lambda_{xx} - \Lambda_{yy})}{\Lambda_{yy}a_x^2  + \Lambda_{xx} a_y^2 + \Lambda_{xx}\Lambda_{yy}\Lambda^{-1}_{zz} a_z^2}\:.
$$

Since the integral $\int (3 u^2 - 1) du$ equals 0, Eq.~(\ref{Jxy}) can be represented in the following equivalent form
\begin{equation} \label{Jxy0}
J =\int\limits_{-1}^1  du \frac{(3u^2 - 1)}{ {\rm e}^{(- \hbar \omega/2 - \mu + b Q u )/T} + 1 } + (\mu \to -\mu)\:.
\end{equation}
At an arbitrary temperature, the integral \eqref{Jxy0} can be calculated analytically, but the corresponding expression (containing polylogarithms) is rather cumbersome, and we do not present it here. Instead, we give the expressions for $J$ at low temperature ($T \to 0$)
\begin{equation} \label{int2}
J = - \theta\left( 1 - |u_0|\right) u_0 (1 - u_0^2)\:,
\end{equation}
and high temperature ($b \hbar \omega/2 \ll T$)
\begin{equation} \label{int3}
J \approx \frac{b^2}{15}  (\hbar \omega)^2  {\cal F}''\:.
\end{equation}
Here, $\theta(x)$ is the Heaviside step function,
 \begin{equation} \label{u0}
u_0 = \frac{1}{b} \left( 1 - \frac{2 |\mu|}{\hbar \omega}\right)\:, \quad {\cal F}(\varepsilon)=f_0(-\varepsilon)-f_0(\varepsilon)\:,
 \end{equation}
the stroke means derivative with respect to $\varepsilon = \hbar \omega / 2$.
\section{Circular photocurrent with account for the cubic nonlinearity}
\label{Sect. V}

In the presence of the cubic terms, the allowance for the tilt is not necessary for getting a photocurrent~\cite{GolubIvch}. Therefore in this section, we set $d_0({\bm k}) \equiv 0$.
We assume nonlinear contributions in the Hamiltonian (\ref{linnonlin}) to be small compared with the linear one. In the first order of nonlinearity $P_i^{(3)}(\bm k)$ the photocurrent~\eqref{gMOmega} can be presented as a sum of three terms
\begin{equation} \label{4contr}
{\bm j} = {\bm j}^{(3)}_v + {\bm j}^{(3)}_{\Omega} + {\bm j}^{(3)}_{\delta} 
\end{equation}
determined by nonlinear corrections respectively to the group velocity ${\bm v}_{{\rm s}\mbox{-}{\rm o}}$, Berry curvature~(\ref{Berrygeneral}) and energy entering $\delta$-function in Eq.~(\ref{MOmega}). In an explicit form, these corrections read
\begin{align} \label{approxn}
d  \approx & Q + \frac{{\bm Q} \cdot {\bm P}^{(3)}}{Q}\:, 
\\  
v_{{\rm s}\mbox{-}{\rm o}, j} =& \frac{2}{\hbar^2 \omega} \left( \beta_{ij}\beta_{il} k_l + \beta_{i j} P^{(3)}_i + \beta_{il} k_l \frac{ \partial P_i^{(3)} }{\partial k_j}\right), \nonumber\\
\Omega_{{\bm k},i} =&  \frac{4}{(\hbar \omega)^3} \biggl[ {\rm det}(\hat{\bm \beta}) k_i + 
{\bm P}^{(3)} \cdot\left( \frac{\partial \bm Q }{\partial k_{i+1}} \times \frac{\partial {\bm Q}}{\partial k_{i+2}}\right)\nonumber
\\
&+  {\bm Q} \cdot \left( \frac{\partial \bm Q }{\partial k_{i+1}} \times \frac{\partial {\bm P}^{(3)}}{\partial k_{i+2}}  -  \frac{\partial \bm Q }{\partial k_{i+2}} \times \frac{\partial {\bm P}^{(3)}}{\partial k_{i+1}} \right)  \biggr] \:,\nonumber
\end{align}
where the value of $d$ in the denominator is replaced by $\hbar \omega / 2$ and, for convenience, we use the variable ${\bm Q}$ defined according to Eq.~(\ref{QLambda}). Next we successively substitute the nonlinear corrections into Eq.~\eqref{gamma_tot}, switch from summation over ${\bm k}$ to integration over ${\bm Q}$, and average the integrand over the directions of the vector${\bm Q}$ according to the rules for tensor integrals~\cite{combinator} 
\begin{align}
\left\langle\frac{Q_{i_1}Q_{i_2}Q_{i_3}Q_{i_4}}{Q^4}\right\rangle = \frac{\delta_{i_1 i_2 } \delta_{i_3 i_4 } + \delta_{i_1 i_3 }\delta_{i_2 i_4 } + \delta_{i_1 i_4 }\delta_{i_2 i_3 }}{15}, \nonumber
\\
\left\langle\frac{Q_{i_1}Q_{i_2}Q_{i_3}Q_{i_4}Q_{i_5}Q_{i_6}}{Q^6}\right\rangle = \frac{1}{105} \left( \delta_{i_1 i_2 } \delta_{i_3 i_4 } \delta_{i_5 i_6 } + \dots\right)\:,
\nonumber
\end{align}
where the ellipsis ... indicates 14 remaining products of three $\delta $-functions with paired indices. The final result for the component $\gamma_{xy}$ is
\begin{equation}
\label{gamma_cub_fin}
\gamma_{xy} = \frac{3N\mathcal{C}\Gamma_0\tau_p}{80} (\hbar\omega)^2 {\cal F}(\hbar\omega/2) \xi_{xy}\:,
\end{equation}
where
\begin{equation} \label{chi}
\xi_{xy} = \sum_{s=x,y,z} (\Xi_{sxy}-\Xi_{syx}),
\end{equation}
\begin{align} \label{Xisxy}
&\Xi_{sxy} = C_{yxkq}\Lambda^{-1}_{kq}+2C_{kxkq}\Lambda^{-1}_{yq}+C_{m\alpha\beta\gamma}\Lambda_{xm}\Lambda^{-1}_{\alpha\beta}\Lambda^{-1}_{\gamma y} \nonumber \\ 
&+4 \left(C_{zzxs}\Lambda^{-1}_{ys}-C_{yzxs}\Lambda^{-1}_{zs}  + C_{xxxs}\Lambda^{-1}_{ys}-C_{yxxs}\Lambda^{-1}_{xs}\right) \:,
\end{align}
and $\Xi_{syx}$ is obtained from (\ref{Xisxy}) by the replacement $x \leftrightarrow y$. 
In the particular case of $\beta_{ij} = \beta \delta_{ij}$, $P_x^{(3)} = C^{(3)}_{xyyy}k_y^3$, $P_y^{(3)} = C^{(3)}_{yxxx}k_x^3$ and $P_z^{(3)} = 0$, we arrive at the result of the 
work~\cite{GolubIvch}
\begin{equation}
\label{gamma_PRB}
\gamma_{xy} = {\cal C}{3N\Gamma_0\tau_p\over 20 \beta^2} (\hbar\omega)^2 {\cal F}(\hbar\omega/2) \left(C_{xyyy}-C_{yxxx}\right)\:,
\end{equation}
where $C_{jlmn} = C^{(3)}_{jlmn}/\beta$.
\section{Discussion}
\begin{figure}[t]
\includegraphics[width=\linewidth]{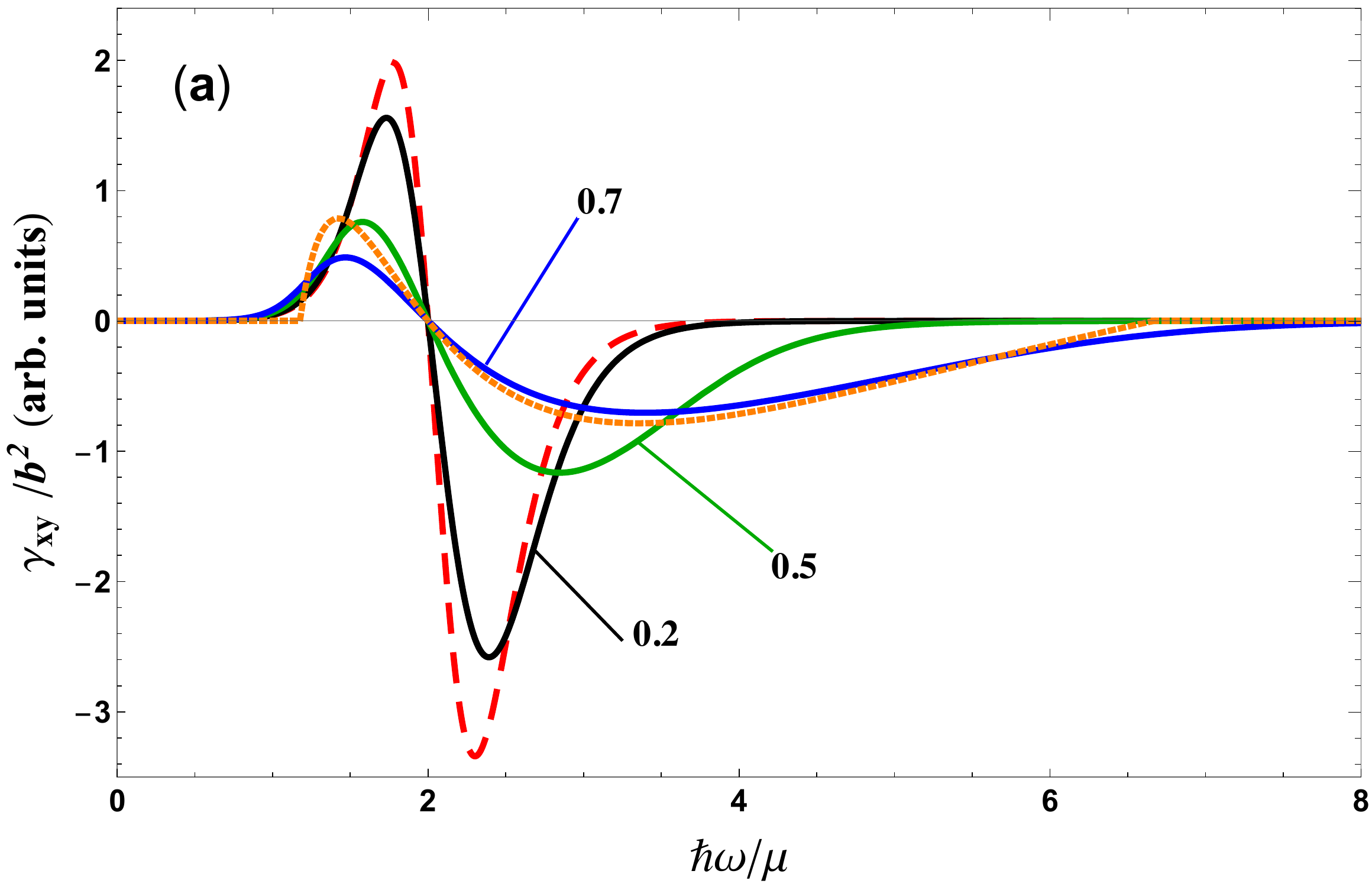}
\includegraphics[width=\linewidth]{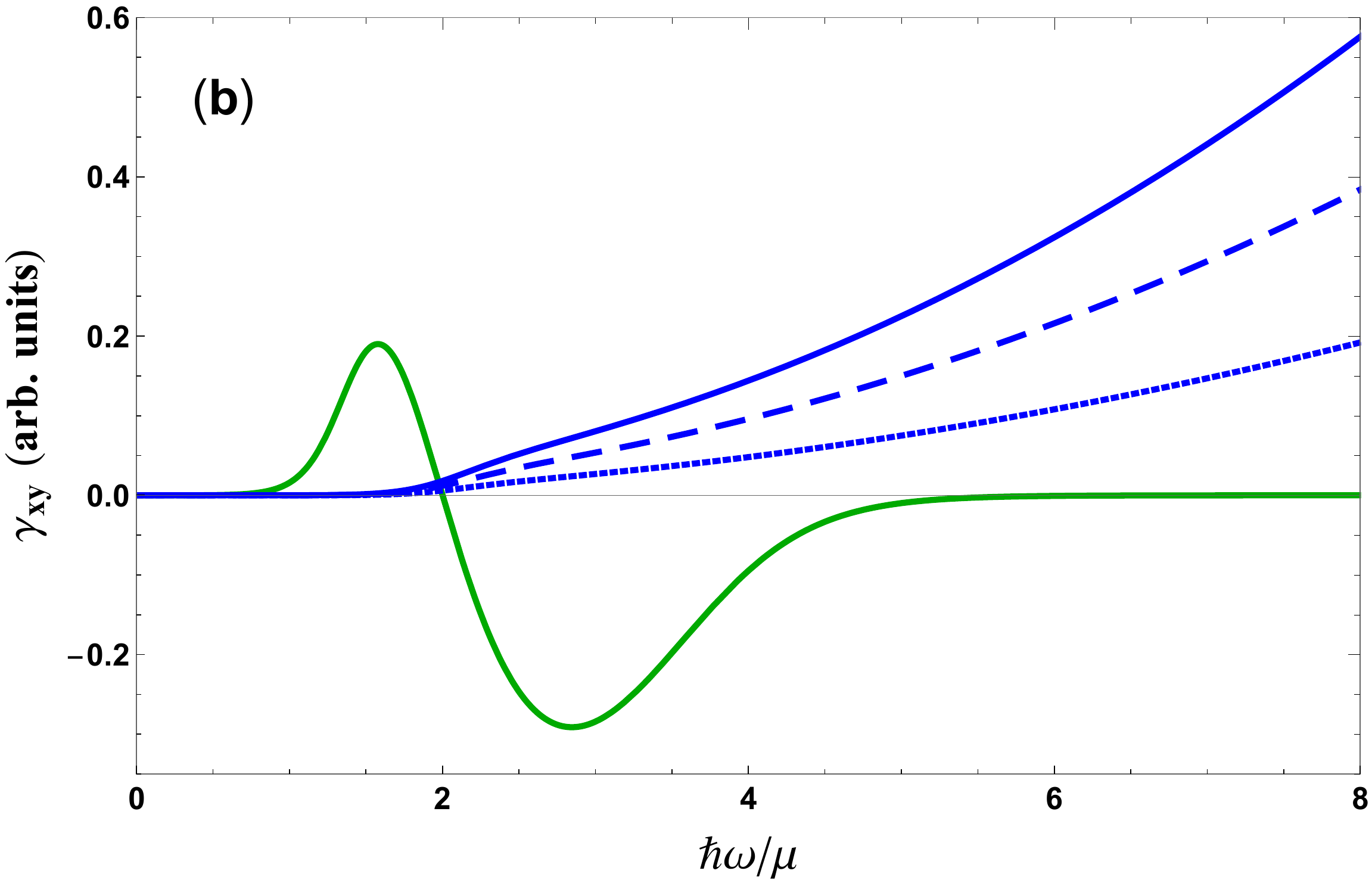}
\caption{
({a}) Photocurrent versus the photon energy in the model with an anisotropic linear spectrum  for 
$T/\mu$~=~0.1. Solid curves show numerical calculation using Eqs.~(\ref{int})-(\ref{Jxy}) for three values of $b = ~ 0.2, 0.5$ and 0.7; the dashed curve is an approximate calculation after Eq. 
(\ref {int3}), and the dotted curve shows the dependence calculated for $T$ = 0 after Eq. (\ref{int2}). ({b}) Comparison of the spectral dependences of the circular photocurrent calculated at $T/\mu = 0.1$ in the model with allowance for the tilt, $b = 0.5$, (green curve) and the model with zero tilt and nonzero cubic nonlinearity: the dotted, dashed and solid curves correspond to $\mu^2 \xi_{xy}/\chi_{xy}=0.03, 0.06$ and 0.09.}
\label{fig:spectrum}
\end{figure}

The dependence of the circular photocurrent on the frequency in the model with an anisotropic linear spectrum is illustrated in Fig.~\ref{fig:spectrum}(a). As can be seen from Eqs.~(\ref{int})--(\ref{Jxy}), the photocurrent depends on two scalar parameters $\chi_{xy}$ and $b =|\hat {\bm \beta}^{-1} \bm a|$. The first parameter is independent of the modulus $|{\bm a}|$, it determines the scale of the photogalvanic effect but not the shape of the photocurrent frequency dependence, and when arbitrary units are chosen along the ordinate axis it falls out. The $b$ parameter affects both the shape and the peak-to-valley displacement of the $\gamma_{xy}(\hbar \omega)$ dependency. Due to the charge symmetry of the spin-dependent part of the Hamiltonian~(\ref{linearH}), the integral $J$ is invariant to the replacement of $\mu$ by $-\mu$, which clearly follows from its representation in the form~(\ref{Jxy0}). For any temperature one of the two contributions to (\ref{Jxy0}) at $\hbar \omega = 2|\mu|$ vanishes, and the other, at $|\mu| \gg T$, is small and can be neglected.

Therefore, the integral $J$ with $\hbar \omega = 2 |\mu| $ should change sign, as is the case for all curves in Fig.~\ref{fig:spectrum}(a). The solid curves show the results of an exact calculation using Eqs.~\eqref{int}--\eqref{Jxy}, the dashed curve corresponds to the calculation using the formula~\eqref{int3}, obtained by expanding the integral $J$ in a small parameter~$b$. According to (\ref {int2}) at zero temperature the photocurrent is generated within the photon energy interval~\cite {GolubIvch}
\begin{equation} \label{range}
\frac{2|\mu|}{1 + b} < \hbar \omega < \frac{2|\mu|}{1 - b}  \:,
\end{equation}
i.e., in the energy window $\Delta = 4 |\mu| b/(1-b^2)$. For $b = 0.7$, this window lies in the range of $1.18 |\mu|$ to $6.7 |\mu|$ in agreement with the dotted curve in Fig.~\ref{fig:spectrum}(a).
It should be noted that the same frequency restriction is also imposed when the tilt and the quadratic nonlinearity ${\bm P}^{(2)}({\bm k})$ in  (\ref{linnonlin}) are taken into account together.
At a finite temperature, the spectral region within which the effect is significant goes beyond the interval (\ref{range}), but in general it is determined by the largest between $ \Delta $ and $ T $. Note also that it follows from Eq.~(\ref {u0}) that, at $ T = 0 $, the photocurrent does not depend separately on $ \hbar \omega $ and $ \mu $ but is a function of the ratio $ \hbar \omega / \mu $, whereas, for $ T \neq 0 $, the photocurrent receives an additional dependence on the ratio $ T / \mu $. 

The frequency dependence of the contribution to the photocurrent arising due to the cubic nonlinearity is fundamentally different from the behavior of the curves in Fig.~\ref{fig:spectrum}(a). Indeed, it follows from Eq.~\eqref{gamma_cub_fin} that $\gamma_{xy} \propto \omega^2 {\cal F}(\hbar \omega/2)$. At $T \ll \hbar \omega$, the population difference ${\cal F}(\hbar\omega/2)$ between the initial and final states is negligible if $\hbar\omega < 2 |\mu|$ and close to unity if the photon energy $\hbar\omega$ exceeds $2 |\mu|$, so that above the absorption threshold $\hbar\omega = 2 |\mu|$, the photocurrent is proportional to the  radiation frequency squared, see~Fig.~\ref{fig:spectrum}(b). For comparison, in this figure we also show the spectral dependence of $\gamma_{xy}$ obtained in the linear model with a tilt.
 
Two experimental articles have been published on the circular photogalvanic effect in the TaAs Weyl semimetal. Kai Sun et al.~\cite {China} measured the circular photocurrent at the photon energies so high ($\hbar\omega \approx 2.38$~ eV) that the optical transitions occurred far from the Weyl points. In the work~\cite{Gedik}, the circular photocurrent was registered under a CO$_2$ laser excitation ($ \hbar \omega = 120 $~meV), and the analysis was performed in the model of the linear effective Hamiltonian~\eqref{linearH}. Our calculation shows, however, that the linear model gives zero photocurrent at high frequencies, and the generation of the photocurrent can be explained by a ${\bm k}$-nonlinear spin-dependent contribution to the effective electron Hamiltonian.
 
\section{Summary}

We have analyzed the relationship between the electron energy dispersion near the Weyl node and the nine components of the $\beta_{ij}$ matrix which determines the effective Hamiltonian in the vicinity of the node. The dispersion is fixed by six linearly independent components of the diagonal positive-definite matrix $\hat{\bm \Lambda} = \hat{\bm \beta}^T \hat{\bm \beta}$, whereas the $\hat{\bm \beta}$ matrix is determined by these six components, the three Euler angles of the orthogonal transformation of a pair of the basis states at the Weyl node and the chirality ${\cal C} = {\rm sgn}\{{\rm det}(\hat{\bm \beta})\}$.

We have developed the theory of the circular photogalvanic effect in the Weyl semimetals with mirror symmetry. The circular photocurrent is calculated for an arbitrary anisotropic Hamiltonian which includes terms both linear and nonlinear in $\bm k$. Unlike a pure spin current, the electric current is expressed in terms of the components of the $\hat{\bm \Lambda}$ matrix and combinations of the products of $\beta_{ij} $ with components of the higher-order matrices that are invariant to the orthogonal transformation of the basis states at the Weyl node. It has been established that the circular photocurrent in semimetals of the $C_{4v}$ symmetry is non-zero in the model with a tilt if its energy spectrum is anisotropic in the plane perpendicular to the $C_4$ axis. The linear model gives a non-zero photocurrent only in a finite frequency range, while an allowance for the $\bm k$-cubic corrections to the Hamiltonian leads to a circular photocurrent that grows quadratically with the frequency in the entire range of direct optical transitions near the Weyl nodes.

\acknowledgements
The work is partially supported by the Russian Foundation for Basic Research (Grant No.
19-02-00095). L.E.G. acknowledges the support by Foundation for the Advancement of Theoretical Physics and Mathematics ``BASIS''. E.L.I. is grateful to the Academy of Finland  (grant No~317920) for support in participation in the International workshop NPO 2018. 

\appendix

\section{Getting one of the solutions of the matrix equation (\ref{Lambda})}
\label{App1}
As a particular matrix $\hat{\bm \beta}_0$ satisfying Eq.~(\ref{betaD}) we take one of the solutions of the equation
\begin{equation} \label{beta2a}
\hat{\bm \beta}^2 = \hat{\bm \Lambda}\:.
\end{equation}
Let us convert the $\hat{\bm \Lambda}$ matrix to a diagonal matrix
\[
\hat{\bm D}_d \hat{\bm \Lambda} \hat{\bm D}_d^{-1} = \left[ 
\begin{array}{ccc}\lambda_1^2 & 0 &0\\0& \lambda_2^2 &0 \\0&0& \lambda_3^2 \end{array} \right]\:,
\]
where $\lambda_1^2, \lambda_2^2, \lambda_3^2$ are eigenvalues of this matrix, and $\hat{\bm D}_d$ is the matrix of the corresponding orthogonal transformation. Then the matrix sought for has the form
\begin{equation} \label{beta2}
\hat{\bm \beta}_0 = \hat{\bm D}_d^{-1} \left[ \begin{array}{ccc}\lambda_1 & 0 &0\\0& \lambda_2 &0 \\0&0& \lambda_3 \end{array} \right] \hat{\bm D}_d\:,
\end{equation}
where, for definiteness, we assume all the three values of $ \lambda_1, \lambda_2$ and $\lambda_3 $ to be positive. The matrix  (\ref{beta2}) can be expressed in terms of the matrix $\hat {\bm \Lambda}$ and its eigenvalues, see e.g. \cite{Franca},
\begin{equation} \label{solution2}
\hat{\bm \beta}_0 = \frac{{\rm I}_{\beta}\sqrt{\Delta_{\Lambda}} \hat{\bm I} + \frac12 \left( {\rm I}_{\beta}^2 +  {\rm I}_{\Lambda} \right)  \hat{\bm \Lambda} - \hat{\bm \Lambda}^2}{\frac12 {\rm I}_{\beta} \left({\rm I}^2_{\beta} - {\rm I}_{\Lambda} \right)- \sqrt{ \Delta_{\Lambda}} } 
\:,
\end{equation}
where $\hat{\bm I}$ is a unit 3$\times$3 matrix, $\Delta_{\Lambda} = {\rm det}(\hat{\bm \Lambda})$, while ${\rm I}_{\Lambda} = \lambda_1^2 + \lambda_2^2 + \lambda_3^2$ and ${\rm I}_{\beta} = \lambda_1 + \lambda_2 + \lambda_3$ are the traces of the matrices $\hat{\bm \Lambda}$ and $\hat{\bm \beta}$. 
Note that although the matrix $\hat{\bm \beta}_0$ is diagonal the set of matrices (\ref{betaD})  satisfying Eq.~(\ref{Lambda}) includes both diagonal and off-diagonal matrices.

The orthogonal transformation matrix in Eq.~(\ref{betaD}) is given by $\hat{\bm D}(\varphi,\theta,\psi) = {\cal C} \hat{\bm \beta} \hat{\bm \beta}_0^{-1}$. One can show that  its transpose is indeed equal to its inverse: $\hat{\bm D}(\varphi,\theta,\psi)^T = \hat{\bm D}(\varphi,\theta,\psi)^{-1}$.  
The Euler angles are defined by the direction of a single rotation about some axis parallel to the unit vector ${\bm n}$ and the angle of this rotation $\Phi$. The vector ${\bm n}$ is an eigenvector of the rotation matrix, i.e. $\hat{\bm D}(\varphi,\theta,\psi) {\bm n} = {\bm n}$.

\end{document}